# Unpacking the "Black Box" of AI in Education

Nabeel Gillani[1*], Rebecca Eynon[2], Catherine Chiabaut[3] and Kelsey Finkel[3]
[1]Massachusetts Institute of Technology, USA // [2]University of Oxford, UK // [3]The Robertson Foundation, USA // ngillani@mit.edu // rebecca.eynon@oii.ox.ac.uk // catherine.chiabaut@robertson.org // kelsey.finkel@robertson.org
[*]Corresponding author

**ABSTRACT:** Recent advances in Artificial Intelligence (AI) have sparked renewed interest in its potential to improve education. However, AI is a loose umbrella term that refers to a collection of methods, capabilities, and limitations—many of which are often not explicitly articulated by researchers, education technology companies, or other AI developers. In this paper, we seek to clarify what "AI" is and the potential it holds to both advance and hamper educational opportunities that may improve the human condition. We offer a basic introduction to different methods and philosophies underpinning AI, discuss recent advances, explore applications to education, and highlight key limitations and risks. We conclude with a set of questions that educationalists may ask as they encounter AI in their research and practice. Our hope is to make often jargon-laden terms and concepts accessible, so that all are equipped to understand, interrogate, and ultimately shape the development of human-centered AI in education.

**Keywords:** K-12 education, Artificial intelligence in education, Educational data mining, Learning analytics, Natural language processing

## 1. Introduction

Rapid advances in artificial intelligence (AI) over the past several years have raised new questions about the role that machines might play in both promoting and impeding humanity. The field of education has been no different. Emerging AI capabilities are enabling machines to fuse and make sense of larger, more diverse datasets in increasingly efficient ways. While these affordances of scale, diversity, and efficiency might help generate insights and guide actions to improve educational opportunities and outcomes, they also come with several technical limitations and related practical risks—like failures to generalize and identify causal relationships—that threaten to perpetuate unfair or harmful applications. Thus, and rightfully so, the re-emergence of AI has sparked new debates about the political, pedagogic, and practical implications of its application in educational contexts (Shum & Luckin, 2019). These debates are critical, especially if we wish for machines to be able to better-serve the human actors—teachers, learners, administrators, and others in education—who may benefit from their emerging capabilities.

Engaging productively in these debates, however, requires one to understand some of the methodological paradigms and practices specific to Artificial Intelligence in Education (AIED). However, researchers and practitioners not trained in computer science or engineering may find the rapidly advancing field of AI inaccessible. In this article, we try to address this gap, providing an overview of the meanings, methods, and limitations of AI as a re-emerging field and how these intersect with AI's applications in education. In doing so, we hope to build on previous introductions to this topic (e.g., Luckin, 2018; Holmes et al., 2019) and critical works that connect data models with ethical and social debates (Perrotta & Williamson, 2018; Perrotta & Selwyn, 2020). By opening up the "Black Box" of AI for those outside of the field, we hope to further human-centered AI in education by empowering all stakeholders, regardless of disciplinary background, to contribute to the development of AI that recognizes and champions human capabilities (Li & Etchemendy, 2018; Yang et al., 2021).

## 2. Defining "AI"

As Artificial Intelligence evolves, the term "AI" has acquired mystical and rhetorical qualities (Eynon & Young, 2020). Some recent advances are impressive: we now have machines that can discover new drug formulas (Popova et al., 2018), predict elusive protein structures (AlphaFold Team, 2020), generate full-length written stories (Brown et al., 2020), and beat world-class performers in games like Starcraft, Go, and Chess (AlphaStar Team, 2019; Silver et al., 2018). Still, demystifying AI is an important first step towards understanding its inner workings and applications. While the capabilities and performance of today's AI systems are unprecedented, many of the core algorithms that govern how they work are rooted in methods dating back to the early 20th century (Tuomi, 2018). Furthermore, while current incarnations of AI have achieved unprecedented degrees of

99

sophistication, the "I" of AI systems remains quite rudimentary—as evidenced by how poorly these systems often perform on tasks that humans find intuitive. Such technical limitations entail important risks and ethical considerations which have significant bearings on the application of AI to the field of education. Before delving into these risks, we expand on two schools of AI that are frequently used in education—machine learning and rule-based AI—and outline some of their common applications.

**2.1. Machine learning-based AI**

*2.1.1. Machine learning paradigms: supervised, unsupervised, and reinforcement learning*

Machine learning algorithms are designed to mine large datasets to uncover—or "learn"—latent rules and patterns that may help inform some future decision. For example, imagine a large school system has asked a research team to develop a tool that accurately predicts what a student's GPA will be at the end of a given school year. "Supervised learning" is one approach to machine learning that could help them tackle this problem. With supervised learning, machines are provided a historical dataset of inputs, or features (e.g., student-level characteristics like demographic data, attendance records, test scores), along with a target output, or attribute (e.g., GPA). A model is then applied to the dataset to learn how these features map to the target attribute by testing out different hypotheses about the relationship to or path from student-level characteristics to GPA. The labels (historical GPAs) of the data in the set help "supervise" the model by indicating how far off its predictions are from the observed or existing (i.e., ground-truth) values. This occurs iteratively for each data point, eventually "training" the model by updating the weights it attaches to the inputs or other variables it uses to make predictions. These weights are often the quantities "learned" by the machine (hence, the term "machine learning"). Linear regression offers a classic approach to supervised machine learning. In fact, many modern approaches using neural networks (described in more detail below), while often described in quasi-mystical terms in press articles, operate in fundamentally similar ways and seek to achieve similar outcomes as linear regression. The scenario in the Supplementary Materials offers additional details about these similarities and differences.

In contrast to supervised learning, "unsupervised learning" is a process by which a machine performs statistical pattern recognition without access to ground-truth labels for the desired output. A common application of unsupervised learning is clustering. Say a school system asked a research team to develop a "typology" of students based on their different characteristics, to help design and target student supports. They could use a standard clustering algorithm (e.g., the popular "k-means" algorithm proposed by Hartigan and Wong, 1979) to learn a grouping of students that differentiates them from other (also automatically inferred) groups. Our resultant groups—or clusters—may comprise students who perform similarly; who take similar classes; live in similar parts of the city; or have some other set of related characteristics.

A third paradigm of machine learning is "reinforcement learning," which has recently been used, among other applications, to develop powerful gameplay systems (e.g., Mnih et al., 2015; Silver et al., 2018). In education, some researchers have started to explore applications of reinforcement learning to intelligent tutoring systems (Reddy et al., 2017). At its core, a reinforcement learning algorithm accepts as an input the state of the world (e.g., the questions a student has answered correctly or incorrectly in an intelligent tutoring environment of a game) and uses this to decide upon some action (e.g., which question to ask the student next). The action—either immediately or over the course of time—eventually contributes to some outcome (e.g., mastering a concept). The value of this outcome is then used to assign positive or negative rewards to the algorithm to encourage or discourage similar actions when faced with similar states of the world in the future. Reinforcement learning algorithms have been around for several decades (Kaelbling et al., 1996), but have resurged over the past few years with large quantities of training data and computational resources more readily available.

*2.1.2. Machine learning philosophies: frequentist and Bayesian*

The paradigms above reflect a "frequentist" philosophy of machine learning: inferences (like predictions, cluster assignments, and other insights that inform decisions) are made largely based on the frequencies of patterns revealed in the training data (Bayyari & Berger, 2004). By contrast, "Bayesian" machine learning models explicitly incorporate pre-existing beliefs ("priors") alongside the patterns revealed by training data to produce some posterior "belief" or inference about the world (Bayyari & Berger, 2004).

Say, for example, a biased coin is tossed 100 times and yields heads on 30 instances. Two friends make a bet that they can infer the true bias of the coin and predict the next 100 tosses. One trains a frequentist machine learning model on the observed coin tosses, which simply factors in the observed data. The model infers the bias as equalling a 30% likelihood of landing on heads. The other friend, however, devises and trains a Bayesian model: in addition to factoring in the observed number of heads, she also factors in a prior belief drawn from most normal coin tossing activities: that there is distribution of possible chances that the coin will land on heads (centered around what we usually expect from coins, 50%). On the next 100 tosses, both observe 40 heads. In this instance, factoring in prior beliefs into the model—instead of simply trusting the observed data—produced an inference of the coin's bias as falling between 30% and 50%, which was more accurate than trusting only the data from the initial set of coin tosses.

Since they rely on both observed data *and* prior beliefs, Bayesian methods can sometimes help overcome sparsity in datasets—like our limited number of coin tosses—in order to make more accurate predictions. In other cases, such prior beliefs may themselves be biased and therefore make models less accurate than if they were trained only on observations. Whether a Bayesian or frequentist model is more appropriate to use depends on the nature of the problem at hand. Interestingly, many believe that the rich structure of Bayesian models reflects aspects of human cognition (Tenenbaum et al., 2011), making them "truer-to-nature" AI. However, many methods for conducting Bayesian posterior inference do not scale well to large datasets, making them difficult to deploy in several real-world settings. In practice, many approaches to machine learning can be implemented from either a Bayesian or Frequentist point of view.

### *2.1.3. The rise of deep learning*

Deep learning—a popular approach to machine learning—has become the dominant school of AI in recent years owing largely to a resurgent interest in neural networks. Neural networks take inspiration from connectionist philosophies of cognitive science (Elman et al., 1996) and generally operate by learning (possibly nonlinear) relationships between several input variables in order to produce predictions as accurately as possible (see the machine learning scenario in the Supplementary Materials for more details). They are the core, modular building blocks that make deep learning systems "deep": combining smaller neural networks together to form larger ones by feeding the outputs of one as inputs to another can enable the discovery of more complex and granular relationships between these inputs and outputs (LeCun et al., 2015). Neural networks can manifest through a number of different algorithmic architectures, e.g., Recurrent Neural Networks (RNNs (Goodfellow et al., 2016)), Convolutional Neural Networks (CNNs (Goodfellow et al., 2016)), and Transformers (Vaswani et al., 2017)—which underpin recent advances in natural language processing like the popular BERT model (Devlin et al., 2019). Each of these architectures differ in how they process and transform inputs into outputs. Furthermore, RNNs and Transformers are generally better-suited for tasks that involve time-series data, whereas CNNs are often applied to image processing problems. Still, while their precise structures and implementations may differ, many of these architectures are trained, evaluated, and eventually used in similar ways.

Deep learning has been driven by advances across three major areas over the past several years: data, algorithms, and hardware. Large, easy-to-access datasets have enabled, for example, the recent "GPT-3" language model—which is trained on over 570 gigabytes of text found across the open internet (Brown et al., 2020). The model is simply trained to predict the next word in a corpus of text given some sequence of preceding words. The result is a powerful system that can generate entire believable stories—an exciting possibility, but also of particular concern in our current era of misinformation (OpenAI Team, 2019).

In cases where large datasets are not available for a specific task, algorithmic advances like "transfer learning" can help (Pan & Yang, 2009). Transfer learning enables a model to "pre-train" itself—i.e., initialize its parameters—using the outputs of a training process conducted for a separate but related task for which enough data *is* available. The model can then "fine-tune" on—or adapt itself to—a smaller dataset that more closely represents the task at hand. For example, early warning systems to detect students likely to drop out may be developed for districts that lack a breadth or depth of historical data by "borrowing" the predictive capacities of models pre-trained on data from larger school settings as a starting point (Coleman et al., 2019). Pre-training, however, may also contribute to the amplification and propagation of biases across models.

Finally, recent hardware accelerations like Graphics Processing Units (GPUs) and Tensor Processing Units (TPUs, Hennessy & Patterson, 2019) are enabling more time-efficient computation, yet their energy demands and associated costs have raised concerns about their potential environmental impacts (García-Martín et al., 2019) and contribution to widening divides between the AI capabilities of large companies and smaller research groups (Hao, 2019).

## 2.2. Rule-based AI

Machine learning systems can be powerful, particularly for problems where the "rules" needed to produce certain outcomes (e.g., the weights to be applied to students' characteristics in order to produce GPA predictions) are not known and hence must be inferred from data. However, there are also problems for which the rules *are* known, but applying these rules can be cumbersome or time-consuming. For these types of problems, "rule-based" approaches to AI—in which computers manipulate data based on a set of pre-defined logical propositions, instead of ones inferred from patterns in the data—are often used.

One such problem in education is school bus routing. Large school districts often have fleets of school buses that must be scheduled and routed to different stops in order to ensure students get to school on time and safely (Bertsimas et al., 2019). In this problem, the rules an AI would consider might include: different carrying capacities for each bus; times by which certain groups of students must get to school; or specific roads buses can and cannot take. A social planner implementing this algorithm may seek to optimize for multiple objective functions: for example, minimizing costs and travel times and/or maximizing the diversity of the student body that travels together on any given bus.

The most naive rule-based AI algorithm would use a brute force approach to solve this problem, evaluating every possible combination of bus, student, and route assignments and selecting the one that yields the most optimal response vis-a-vis our multiple objectives. For many large real-world problems, however, this approach is infeasible and could quite literally take hundreds of years (or longer) to compute (Cook, 2012). To this end, rule-based AI algorithms often use sophisticated solution strategies to prune down a large set of possible combinations to a feasible subset that is much easier and more efficient to search through (e.g., Van Hentenryck & Michel, 2009). Unlike machine learning systems, rule-based models will not necessarily make more accurate decisions with a larger scale or diversity of data. In fact, scale and diversity of data can pose challenges to rule-based AI algorithms because they increase the size and complexity of the problem at hand. This said, these challenges will likely be alleviated by the increased algorithmic and hardware efficiencies afforded by the current wave of AI described above.

While their underlying mechanisms might differ, rule-based AI need not be completely distinct from machine learning. For example, we may have historical data on bus routes and road conditions (e.g., traffic patterns) which we can use to predict travel times. We can then leverage these predicted travel times as inputs into our objective function during the optimization process.

# 3. Applications of AI in education

Despite recent interest in applications of AI and education, the two fields have intersected for some time (e.g., Aleven & Koedinger, 2002)—which has long raised important philosophical and ethical questions. This next section provides an overview of recent applications of AI in education and highlights some of their limitations and broader implications. These examples, far from exhaustive, have been selected in order to highlight the ways in which the scale and diversity of available data—along with improvements in computational efficiency—have created new opportunities for using AI to potentially improve the human condition through educational applications. For a more in-depth review of how AI and other data mining techniques *can* be applied to education-related problems, we refer readers to several existing review papers (e.g., Romero & Ventura, 2010; Koedinger et al., 2015; Fischer et al., 2020).

## 3.1. Intelligent tutoring systems

Intelligent tutoring systems (ITS) are a popular application of AI in education. ITS are tools that seek to adapt to students' existing knowledge and skills, or learning states, to help them build skills in more personalized ways. The "I" in ITS often has different definitions for different tools. For example, some ITS are machine learning-based systems that seek to develop (sometimes Bayesian) learner models trained to maximize the likelihood of a student answering a provided question correctly, conditional on their history of responses (Ritter, 2007). In other cases, developers might simply train a system to predict the likelihood of "correctness" as accurately as possible (e.g., using deep reinforcement learning a la Reddy et al., 2017). These systems then provide students with problems that are most likely to be at their "learning edge"—i.e., the problems they haven't yet answered that they are most likely to answer correctly, given their prior history of answers. These machine learning systems have the capacity to make more accurate predictions of a student's learning edge as they draw on larger and more disparate historical sources of student performance and behavior—many of which are becoming more ubiquitous through computer-aided tutoring and assessment platforms. Other ITS, like (Kelly et al., 2013), pre-define simple rules—like correctly answering three similar questions in a row—to determine if and when a student has mastered some concept.

Experimental evidence has largely shown ITS to be effective in increasing students' grades and test scores (J-PAL Evidence Review, 2019). Of course, grades and test scores offer only one (limited) view into student learning. Crucially, much of the existing efficacy research on ITS has not specifically analysed which underlying AI methods make them more or less effective. As such, it is unclear to what extent machine learning vs rule-based systems are responsible for helping students improve their outcomes. As machine learning technologies continue to offer new opportunities for personalizing instruction, it will be important to identify the precise elements of these systems that offer the greatest promise for enhancing student learning. There is also a need to better understand the contexts in which these ITS systems can be meaningfully deployed as a resource for teachers and students in ways that do not inadvertently narrow the aims and purposes of Education (Biesta, 2015).

### 3.2. Assessment and feedback

Proponents of AI, particularly machine-learning based systems that seek to infer students' knowledge states from the growing scale and diversity of data available on digital learning platforms like Khan Academy, argue these systems have the capacity to obviate the need for explicit formative and summative assessments, by seeking to infer students' knowledge states from the growing scale and diversity of data available on such digital learning platforms (Piech et al., 2015) and other systems instrumented for "learning analytics" (Gašević et al., 2015). After all, if it is possible to know what a student knows based on how they answer questions in an ITS, why administer an assessment at all? This line of reasoning, of course, does not consider the positive effects exam preparation and studying can have on learning (Karpicke & Roediger III, 2008).

Automated assessment of writing submissions is a popular, albeit complex, example of how machine learning might support assessment. To date, most research has focused on training machine learning models to assess foundational attributes of writing—for example, spelling, vocabulary, and grammar. Other systems have used machine learning to train models that are able to replicate human scores for a given essay (Dong et al., 2017). Growing as a writer, however, requires much more than feedback on the mechanics of writing or collapsing a rich composition down to a single grade. To this end, (Fiacco et al., 2019) recently designed a neural network-based machine learning system to identify which rhetorical structures were present in sentences contained within a corpus of research study articles: for example, which sentences sought to describe the study, provide context on the study's methods, or frame new knowledge.

Despite the advancing capabilities of these systems, however, some concerns remain. For instance, it would be important to train these AI on a diverse set of linguistic data to fuel their accuracy and minimize bias. More work also needs to be done to understand how they might inadvertently negatively impact writing development and written work in the same ways as plagiarism detection software has (Ross & Macleod, 2020), and more generally, how student surveillance via constant data collection may impact students (Eynon, 2013). Thus, although assessment and feedback is a core focus of AIED, the most appropriate ways to deploy AI for particular activities and in specific contexts remains an area of debate.

### 3.3. Coaching and counselling

The role of coaches and counsellors in schools are multifaceted, time-intensive, and costly. Researchers have therefore started to explore how some of their tasks can be automated. For example, several studies have demonstrated how text-message reminders can help facilitate specific outcomes normally under counsellors' purview: e.g., ensuring that graduated high school seniors take the steps needed to matriculate at college in the fall (Castleman & Page, 2015) and keeping parents updated about their children's academics (Bergman & Chen, 2019).

Recent efforts have also leveraged AI to enable a richer set of interactions between students and "counselors." A recent study (Page & Gehlbach, 2017) deployed an AI chatbot to answer questions about forms students would need to fill out before starting college at Georgia State University (GSU). The authors indicate that the chatbot was trained using deep reinforcement learning—the same technology that has enabled state-of-the-art advances in automated gameplay (AlphaStar Team, 2019)—though the exact methods for training and evaluating these models in the context of the chatbot are unclear. The researchers found that the AI-powered system was comparable in enhancing college enrolment rates to prior studies that primarily involved human counselors. As more dialogue agent systems are deployed across campuses, the scale and breadth of available linguistic corpora for training models with smarter response strategies are likely to grow. Nevertheless, a number of open research questions persist—particularly concerning how well these systems can serve a diverse student body in answering complex educational questions.

### 3.4. (Large) school systems-level processes

At the school systems-level, AI is being used to achieve several objectives, including the equitable implementation of school choice. Over the past two decades, a strand of economics research has focused on developing rule-based AI algorithms for districts that offer families choices on where to send their children to school. These algorithms have been designed to be "strategy-proof," matching students to schools in ways that do not enable families to "game the system" by mis-stating preferences in order to exploit loopholes that would increase their likelihood of receiving a spot at one of their top choice schools (Pathak & Sönmez, 2008). This is particularly important for those parents who do not have the resources, social capital, or knowledge necessary to "game the system." Of course, "strategy-proofness" only helps further equity to the extent that other parts of the system are also equitable (Goldstein, 2019).

AI has also been used to help with a range of planning and forecasting tasks, particularly in larger school-systems or by those working across large systems of schools. Working with Boston Public Schools, researchers built a machine learning model that forecasts changes in demand for schools in response to certain school choice policy changes (Pathak & Shi, 2015). As more data accrues across the diverse spectrum of families in these systems, such models have the potential to become more accurate—and perhaps also shed more light on the preferences of families who belong to traditionally underrepresented segments of the population. School districts have also turned to rule-based AI systems to help achieve greater logistical efficiency—for example, by producing "optimal" bus routes as discussed earlier in this paper—and to save money (Bertsimas et al., 2019). Yet such systems have been met with mixed reception from some of the families they ultimately impact (Scharfenberg, 2018). Additionally, to improve teacher placement in schools, Teach for America (TFA) designed and tested a matching algorithm similar to the school choice matching algorithm described above, to factor in both teacher and school preferences; TFA subsequently saw a slight positive effect on students' academic outcomes (Davis, 2017). With continued increases in computational efficiency, these rule-based systems promise to be able to operate on larger, more complex problems concerning more students, teachers, and other stakeholders in the years ahead. Yet these need to be developed with an awareness of concerns about the use of such market-driven principles to develop an equitable education system (e.g., Ball, 2017; Biesta, 2015).

### 3.5. Predicting outcomes

Machine learning systems have garnered significant attention for their ability to "predict the future"—often in the form of "early warning systems." These systems, often using different forms of regression, mine large troves of historical student data to predict which students are most at risk of failing an exam, dropping out of high school or college, etc. (Faria et al., 2017). Experimental evidence has suggested that deploying these systems can help reduce chronic absenteeism and course failure (Faria et al., 2017). While early warning systems do not always require machine learning—e.g., a simple rule-based system could trigger a warning if a student's GPA falls below a certain level—machine learning-based systems have the potential to identify and exploit patterns of which school leaders may not be aware. These systems can also pool data across disparate contexts to improve individual predictions. For example, small school districts might face a "cold start" problem: they simply do not have enough historical data to train an accurate machine learning model—requiring them to "borrow" data from other school districts to improve accuracy (e.g., Coleman et al., 2019). Increasing scale and diversity of data may enable such applications of transfer learning, and more generally, extend the possible applications of machine learning to educational settings that have previously been left out.

Unfortunately, these warning systems can have several drawbacks. Being able to predict how well a student is going to do in a particular class might help encourage students to take more advanced classes (Bergman et al., 2021)—but it could also lead to tracking, which might limit a student's desire and ability to explore new topics, particularly in college and university. School leaders may also struggle to calibrate interventions based on the outputs of a model. If a model indicates the probability that any given student drops out of high school, at what point should an intervention be triggered—when there is a 20% chance of a student dropping out? 51%? 90%? Even if a school leader feels equipped to intervene after analyzing the data, there is a fundamental question about the obligation to act (Eynon, 2013; Prinsloo et al., 2017; Hakimi et al., 2021): which students should receive support? And what if the model has a high false-negative rate—meaning there could be many students who actually need intervention but weren't flagged by the model as such? These are difficult questions and, at present, there are no standardized responses; school systems approach these questions differently depending on their own knowledge and needs.

## 4. Limitations and risks of modern machine learning systems

Readers from varied sub-fields of education, learning sciences and data science will bring different critical lenses to the areas and applications previously discussed. Here, we will draw from (Lake et al., 2016) and other researchers to discuss several technical limitations of modern machine learning systems and some risks that arise from them. We will also look at the key gaps that still exist between what many believe AI can do in 2021, what it can actually do (and not do), and how these limitations have important implications for education.

### 4.1. Limitations of modern machine learning systems

*4.1.1. Transparency and interpretability*

Neural network approaches to machine learning are powerful, but their inner workings are usually not transparent, making them difficult to interpret. One implication of this is that it may not be clear which inputs were responsible for driving decisions. For example, in the case of early warning systems, a school leader might be informed of the likelihood of any given student failing a course, but not which characteristics of the student are most associated with this prediction. The school leader might obviate this problem by opting for a more interpretable, non-deep learning-based model, but this may require sacrificing some degree of predictive accuracy. These are not always salient tradeoffs, but when they arise, it is often unclear how they should be made. Fortunately, model interpretability is an active area of deep learning research with several recent advances (e.g., Sundarajan et al., 2017; Kim et al., 2018). These advances are critical for equity and inclusion in education, as they open the door to enabling a wider range of stakeholders—including parents and students who may be affected by such algorithms—to understand, interrogate, and ultimately improve their applications (although see Ananny & Crawford, 2018; Tsai et al., 2019 for discussions of questions of the burden such moves could place on individuals).

A more fundamental issue with machine-learning based systems, even those that do not leverage deep neural networks, is causal attribution. Machine learning models are designed to identify and exploit correlations (not necessarily causal relationships) between variables in order to make predictions. For example, a school leader's early warning system might highlight poverty status, prior grade history, and disciplinary actions as student-level factors associated with a higher likelihood of course failure, without explaining the underlying *causes* of failure. Misunderstanding underlying causes may lead to faulty or incomplete interventions, and ultimately, a perpetuation (or exacerbation) of the underlying educational challenges educationalists are seeking to address. Advances in machine learning methods for causal analysis (e.g., Johansson et al., 2016) are attempting to help separate out correlation from causation. However, grasping a rich understanding of causal processes in settings as complex as education usually requires much more than technical solutions.

*4.1.2. Abstract reasoning and learning how to learn*

Humans are very good at two things that AI-powered machines are not: abstract reasoning and learning how to learn. For example, while machines can learn to play a variety of games better than champion-calibre players, they require training on simulations of hundreds of thousands or millions of games to learn how to do so. Humans, by contrast, often learn gameplay simply by watching someone else play for a few minutes (Lake et al., 2016). This is partly because we are remarkably adept at abstract reasoning: ascertaining the fundamental rules of a particular task to generalize and apply these rules to other similar but distinct endeavors. Teachers do this all the time: unlike most intelligent tutoring systems, they do not need to observe a large number of question responses from a student in order to identify and begin addressing key conceptual gaps.

An important type of abstract reasoning is learning how to learn. Throughout our lives, we have likely played several games, and these experiences have made it easier for us to learn the rules and dynamics of new games. Such "meta learning" is a popular area of machine learning research. At present, however, the complex reasoning done by humans is broken down into discrete processes for the machines, including teaching a machine "where to focus" in the space of input data (Xu et al., 2015) or how to automatically update different parts of its own architecture (Andrychowicz et al., 2016; Zoph & Le, 2016) in order to make better predictions. Perhaps unsurprisingly, this machine "meta learning" generally lacks the higher-order thinking, reflection, and planning woven throughout human meta-learning. Without this ability to learn how to learn, we must be skeptical of how well AI can support students, understanding the complex in- and out-of-school factors that impact learning. AI, for example, may be able to suggest problems to students to work on, but will be limited in identifying why students continue to get certain types of problems right or wrong—especially if those factors transcend cognitive, skill-based challenges and extend to the home environment or other social forces affecting the child.

**4.2. Risks that stem from machine learning's limitations**

*4.2.1. Failures in generalizing*

Because machine learning models often fail to develop a deep, intuitive understanding of the task they are built to perform, they can subsequently fail to generalize to new settings than what they were trained for (Murphy, 2012). This sometimes leads them to "catastrophically forget" how to perform tasks (Kirkpatrick et al., 2017), or become brittle in the face of "adversarial" inputs. Adversarial inputs are data examples—often derived by making small perturbations to training set examples—that are designed to fool a machine learning model into making an incorrect decision. As an example, (Brown et al., 2017) showed how an object recognition system that could classify an image as containing a banana with high confidence could easily be fooled into making an incorrect classification simply by adding a small sticker of a toaster to the image. One of the key reasons for this brittleness is probably the fact that the model has not "really learned" what a banana is, beyond a collection of pixels arranged in a certain way. We can play such a scenario out to imagine several concerning possibilities in education, for example: a ranking system that places a student in a remedial class because of their test score similarity to a historical batch of remedial students, without factoring in other variables that might better-indicate their likelihood of succeeding in more advanced courses (Bergman et al., 2021); facial recognition misclassifications in criminal justice applications that lead to the wrongful incarceration of students or their family members (Hill, 2020); and many more. These scenarios have inspired new directions for building more robust deep learning models (e.g., Tjeng et al., 2019), but the need for awareness about what such models are "doing" in technical terms will remain crucial.

*4.2.2. Bias and fairness*

Lacking a general understanding of the "how" and "why" behind most decisions, many machine learning models often recapitulate biases in their training data—and hence, risk perpetuating these biases at scale. For example, a recent study illustrated the drastically poor performance of several commercial facial recognition technologies when seeking to identify the faces of black women—due in part to underrepresentation in their training data (Buolamwini & Gebru, 2018). In healthcare, a system for neurological disorder screening based on human speech proved more accurate for individuals who spoke a particular dialect of Canadian English (Gershgorn, 2018). Such shortcomings prevail in education too—with AIED applications favouring certain groups in the content taught, the ways material is covered, and the accuracy of predictions and appropriateness of interventions (Mayfield et. al., 2019). The UK's intention of using predictive models to assign final grades in the wake of 2020 school closures due to the COVID-19 pandemic illustrates this risk: under the modelling scheme, which was eventually dropped, highly-qualified and capable students from historically lower-performing (and lower-resourced) schools were more likely to receive marks lower than what their teachers would have assigned, whereas students in traditionally high-performing private received higher predicted scores (BBC, 2020).

Ultimately, fairness is a highly complex concept, particularly when applied to education (Mayfield et al., 2019); and when and how educationalists choose to use AIED is itself a complex ethical question, even if and when those AIs are optimized to root out bias. Addressing the technical limitations of machine learning will help mitigate the risks outlined above, but it will be insufficient to preempt the full range of educational and ethical issues related to AIED, specifically the application of AI in practice. Multiple significant and important critiques of AIED, and of the use of data in education more broadly, center on issues such as privacy, instrumentalism, surveillance, performance, and governance (Jarke & Berieter, 2019; Holmes, et al., 2021; Williamson 2017).

We hope that the technical explanations and considerations outlined in this paper can help inform conversations and decision-making around issues of fair and equitable use—even if they are insufficient to resolve them.

## 5. Discussion and conclusion

As we have explored, AI is not "one thing"; in this paper we have focused on the more technical aspects of AI to highlight the myriad of (sometimes complementary) computational techniques that collectively constitute AI. Understanding the workings, limitations, and risks associated with each—and especially those powered by machine learning—is critical to developing and deploying them wisely, thoughtfully, and with proper human oversight. Educationalists who do not have a background in computer science or engineering have a vital role to play in this endeavor. To aid with this, we offer the following guiding questions that educationalists may ask as they encounter applications of AI in education, to ensure AI is used ethically, responsibly, and ultimately to improve the human condition:

- **What kind of AI is it?** The examples contained in this paper illustrate how different types of AI can (and cannot) help solve different problems in education, and may help educationalists form a judgement about their applicability and risks within their own contexts. Asking this question may encourage a recognition of both human expertise and the realities of the 'intelligence' of AI systems.

- **Does the AI enable something that would be difficult or impossible to achieve without it?** Unpacking any benefits of the scale or diversity of data that the AI operates on, or any efficiencies it enables and weighing them against associated risks or limitations may help justify its usefulness. If an AI-powered system does not enable capabilities or benefits that could be achieved without it, it may not be worth deploying. Just because AI *can* be used to power an education technology system, does not mean it *should* be.

- **What are the potential risks or drawbacks of deploying this technology?** Even in cases where AI might enable high-impact new capabilities, there are likely to be critical failure modes that could lead to unintended, perverse outcomes. Understanding the possibility of, and anticipating, these outcomes is of essential importance.

- **How equitably are the anticipated benefits and risks distributed across different groups of students and families?** AI, especially machine learning-based systems, can "learn," replicate, and scale bias and inequity. It is therefore important to question whether AI systems might underserve or discriminate against students and families from low-income or minority backgrounds; with disabilities; experiencing varying levels of linguistic proficiency; or facing other vulnerabilities. Asking about past performance or evidence of bias, or about steps taken to ensure equity in application, could be helpful.

- **If you could wave a magic wand and change anything about this technology, what would it be?** All technologies (including those powered by AI) have been designed with a set of values, practices, and use-cases in mind—and therefore, can be changed, even if they appear opaque or difficult to understand. Those who are closest to the application of AI in educational settings should refuse to accept the status quo, using their observations and wisdom to share feedback with system developers in order to spark changes that help improve the human experience with education.

If and how AI should be designed and used in education remains an active question, which can only be answered through conversations between and across different academic communities. As prior work argues, this will require AI researchers and engineers to work with educationalists to better-understand the theory and practice of education. However, we hope we have successfully argued that equally important is the need for educationalists to understand the more technical aspects of theory and practice of AI, especially when critiquing, rejecting or adapting it for their own efforts. Through the provision of an overview of current AI techniques, their use in education, and key limitations and risks, we hope this article will contribute to these on-going conversations and help advance the quest for AIED to improve the human condition.

## Acknowledgement

We would like to thank John Hood and Eric Chu for many helpful ideas and comments that helped shape this paper. No human subjects were involved in this research. The authors report no conflicts of interest.